\documentclass[prd,twocolumn,showpacs,floatfix,amsmath,nofootinbib,amssymb,floatfix]{revtex4}
\usepackage{graphicx,color,dcolumn,booktabs,bm,multirow}
\usepackage{longtable,lscape}
\usepackage{txfonts}
\usepackage{overpic}
\usepackage{amssymb}
\usepackage{array}
\usepackage{indentfirst}
\usepackage{feynmf}   
\usepackage{slashed}  
\usepackage{cases}
\usepackage{color}
\usepackage{multirow}
\usepackage{epstopdf}
\usepackage{graphicx,color,dcolumn,booktabs,bm}
\usepackage[colorlinks,
            citecolor=blue,
            anchorcolor=red,
            menucolor=red,
            linkcolor=red,
            filecolor=red,
            runcolor=red,
            urlcolor=blue,
            frenchlinks=red]{hyperref}

\def\zcs{Z_{cs}^{[D_s^{(\ast)}\bar{D}^{(\ast)} ]}}
\def\xss{X^{[D_s^{(\ast)}\bar{D}_s^{(\ast)}]}}
\begin{document}

\title{Hunting for the hidden-charm molecular states with strange quark in $B$ and $B_s$ decays}
\author{Qi Wu}
\affiliation{School of Physics and Center of High Energy Physics, Peking University, Beijing 100871, China}
\author{Yan-Ke Chen}
\affiliation{School of Physics and Center of High Energy Physics, Peking University, Beijing 100871, China}
\author{Gang Li}
\affiliation{College of Physics and Engineering, Qufu Normal University, Qufu 273165, China}
\author{Shi-Dong Liu}\email{liusd@qfnu.edu.cn}
\affiliation{College of Physics and Engineering, Qufu Normal University, Qufu 273165, China}
\author{Dian-Yong Chen}\email{chendy@seu.edu.cn}
\affiliation{School of Physics, Southeast University, Nanjing 210094, China}
\affiliation{Lanzhou Center for Theoretical Physics, Lanzhou University, Lanzhou 730000, China}

\begin{abstract}
In the present work, we investigate the productions of the molecular states composed of $D^{(*)}_s \bar{D}^{(*)}$ and $D^{(*)}_s \bar{D}^{(*)}_s$ in the $B$ and $B_s$ decays by using an effective Lagrangian approach. The branching ratios in terms of the model parameter $\alpha$ and the binding energy $\Delta E$ are estimated. Our estimations indicate that the branching fractions are of the order of $10^{-4}$ and the relative ratios are very weakly dependent on the model parameter $\alpha$ and the binding energy $\Delta E$. The predicted ratios are helpful for searching the hidden-charm molecular states with strange quark in the future experiments at Belle II and LHCb.
\end{abstract}

\date{\today}
\maketitle

\section{Introduction}
\label{sec:introduction}

The quark model has achieved great success in classifying the observed hadrons into the mesons and baryons composed of $q\bar{q}$ and $qqq$, respectively~\cite{Gell-Mann:1964ewy,Zweig:1964ruk}. However, a large number of the so-called exotic hadron states have been reported by the Belle, BESIII, CDF, and LHCb Collaborations in recent two decades. Among them, multiquark states with a pair of heavy-antiheavy quarks, also known as the \textit{XYZ} states, are the long-sought goal in experiments and have become excellent candidates for exotic hadron states (see Refs.~\cite{Chen:2016qju,Hosaka:2016pey,Lebed:2016hpi,Esposito:2016noz,Guo:2017jvc,Ali:2017jda,Olsen:2017bmm,Karliner:2017qhf,Yuan:2018inv,Dong:2017gaw,Liu:2019zoy, Chen:2022asf, Meng:2022ozq} for recent reviews).

It is noteworthy that the most observed \textit{XYZ} states are located near the thresholds of a pair of heavy-antiheavy hadrons. The near-threshold nature of the \textit{XYZ} states may shed light on identify their inner structures. For example, the observed masses of the $X(3872)$~\cite{Choi:2003ue} and $Z_c(3900)$~\cite{Ablikim:2013mio, Liu:2013dau,Ablikim:2013xfr} are both near the threshold of $D\bar{D}^*+c.c.$, and the $Z_c(4020)$ is close to the threshold of $D^* \bar{D}^*$~\cite{Ablikim:2013wzq, Ablikim:2013emm}, indicating that they are good candidates of the hadronic molecular state composed of $D^{(*)}\bar{D}^*$. The masses~\cite{Wilbring:2013cha,Khemchandani:2013iwa,Guo:2013sya,Chen:2013omd,Duan:2021pll}, decay properties~\cite{Goerke:2016hxf,Esposito:2014hsa,Dong:2013iqa,Ke:2013gia,Wang:2013cya,Li:2014pfa}, and production processes~\cite{Lin:2013mka,Chen:2016byt,Wu:2019vbk,Liu:2021ojf} of $Z_c(3900)$ and $Z_c(4020)$ have been investigate extensively in the hadronic molecular scenario.

Upon the observations of the hidden-charm states, it is natural to search their possible strange partners, which should be in the vicinity of the thresholds of $D_s^{(\ast)}D^{(\ast)}$, and such kind of hidden-charm structures with strangeness have been predicted in the literatures~\cite{Ebert:2008kb,Lee:2008uy,Dias:2013qga,Chen:2013wca}. Considering the components of $Z_{cs}$, we use $Z_{cs}^{[D_s^{(\ast)} D^{(\ast)}]}$ to represent the possible molecular states composed of $D_s^{(\ast)}$ and $\bar{D}^{(\ast)}$. In view of the fact that the $X(3872)$ and $Z_c(3900)$ were observed in the processes $B\rightarrow K\pi^+ \pi^- J/\psi$ and $e^+ e^-\rightarrow\pi^+ \pi^- J/\psi/\pi^\pm(D\bar{D}^*)^\mp$~\cite{Choi:2003ue,Ablikim:2013mio, Liu:2013dau,Ablikim:2013xfr}, the processes $B\rightarrow K \phi J/\psi$ and $e^+ e^-\rightarrow K^+ K^- J/\psi/K^+(D^-_s D^{*0}+D^{*-}_s D^0)$ are hence suitable for searching for the hidden-charm states with strangeness.  Recently, the BESIII Collaboration observed the $Z_{cs}(3985)$ in the $K^+$ recoil-mass spectrum of the process $e^+ e^-\rightarrow K^+ (D^-_s D^{\ast0}+D^{\ast-}_s D^0)$~\cite{BESIII:2020qkh}, which establishes the first candidate of the charged hidden-charm states with strangeness. The observed mass of $Z_{cs}(3985)$ is close to the threshold of $D_s^{\ast} \bar{D}$, which could be the strange partner of $Z_c(3900)$. Later on, the LHCb Collaboration reported another hidden-charm states with strangeness, $Z_{cs}^+(4000)$, in the $J/\psi K$ invariant mass spectrum of $B^+\rightarrow J/\psi\phi K^+$ process~\cite{LHCb:2021uow}. The observed mass of $Z_{cs}(4000)$ is consistent with the one of $Z_{cs}(3985)$, but the widths are much different. Just as $Z_c(3900)$ explained as $D\bar{D}^*$ molecular state since its near-threshold nature, $Z_{cs}(3985)$ can be naturally explained as $D_s\bar{D}^*$ molecular state~\cite{Meng:2020ihj,Yang:2020nrt,Sun:2020hjw,Wang:2020rcx,Wang:2020htx,Dong:2020hxe,Xu:2020evn,Liu:2020nge,Chen:2020yvq,Ozdem:2021yvo,Yan:2021tcp,Wu:2021ezz}. In addition, the BESIII Collaboration reported their search for the heavier partner of the $Z_{cs}(3985)$ state in the $e^+e^- \to K^+ D_s^{\ast-} D^{\ast0}$, an excess of $Z_{cs}^{\prime }(4120)\to D_s^{\ast -} D^{\ast 0}$ candidates was observed with a significance of $2.1~\sigma$, and the mass of $Z_{cs}^{\prime }(4120)$ was reported to be $(4123.5 \pm 0.7_{\mathrm{stat.}} \pm 4.7_{\mathrm{syst.}})$ MeV~\cite{BESIII:2022vxd}, which is close to the threshold of $D_s^{\ast -} D^{\ast 0}$.

\begin{table*}[htb]
	\caption{\label{tab:be}Theoretical predictions of the masses and binding energies of $D^{(*)}_s \bar{D}^{(*)}_s$ and $D^{(*)}_s \bar{D}^{(*)}$ molecular states. For comparison, \ the threshold of $D^{(*)}_s \bar{D}^{(*)}_s$ and $D^{(*)}_s \bar{D}^{(*)}$ are also listed.}
\renewcommand\arraystretch{1.5}
		\begin{tabular}{p{1.5cm}<\centering p{1.3cm}<\centering p{1.3cm}<\centering p{2.1cm}<\centering p{2.3cm}<\centering p{1.3cm}<\centering p{1.3cm}<\centering p{1.2cm}<\centering p{1.3cm}<\centering p{1.2cm}<\centering p{1.3cm}<\centering}
		\toprule[1pt]			
		System & \multicolumn{2}{c}{$D_{s}\bar{D}$} & \multicolumn{2}{c}{$D^*_{s}\bar{D}^*$} & \multicolumn{2}{c}{$D_{s}\bar{D}_s$} & \multicolumn{2}{c}{$D_{s}\bar{D}^*_s$} & \multicolumn{2}{c}{$D^*_{s}\bar{D}^*_s$}\\
			Threshold    &\multicolumn{2}{c}{3838.1} & \multicolumn{2}{c}{4119.1} & \multicolumn{2}{c}{3937.0} & \multicolumn{2}{c}{4080.8} & \multicolumn{2}{c}{4224.6}\\
			\midrule[1pt]
			Reference & $\Delta E$ & $M$ & $\Delta E$ & $M$ & $\Delta E$ & $M$ & $\Delta E$ & $M$ & $\Delta E$ & $M$\\
			\midrule[1pt]
			\multirow{2}{*}{Ref.~\cite{Hidalgo-Duque:2012rqv}}
			& $2.3^{+7.3}_{-2.3}$ & $3835.8^{+2.3}_{-7.3}$ & $28^{+22}_{-19}$ & $4091^{+19}_{-22}$ & $13^{+13}_{-10}$ & $3924^{+10}_{-13}$ & $46^{+25}_{-23}$ & $4035^{+23}_{-25}$ &
			$48^{+25}_{-23}$ & $4177^{+23}_{-25}$ \\
			& $0.4^{+8.1}_{-0.4}$ & $3837.7^{+0.4}_{-8.1}$ & $22^{+33}_{-24}$ & $4097^{+24}_{-33}$ & $9^{+19}_{-9}$ & $3928^{+9}_{-19}$ & $41^{+39}_{-33}$ & $4040^{+33}_{-39}$ &
			$45^{+40}_{-35}$ & $4180^{+35}_{-40}$ \\
			Ref.~\cite{Wang:2020htx} & ... & ... & $-5.1^{+3.7}_{-5.6}$ & $4124.2^{+5.6}_{-3.7}$ & ... & ... & ... & ... & ... & ...\\
			\multirow{2}{*}{Ref.~\cite{Yang:2020nrt}}
			& ... & ... & $27^{+21}_{-26}/-10\pm4$ & $4092^{+26}_{-21}/4129\pm4$ & ... & ... & ... & ... & ... & ... \\
			& ... & ... & $36^{+24}_{-35}/-19\pm6$ & $4083^{+35}_{-24}/4138\pm6$ & ... & ... & ... & ... & ... & ... \\
			Ref.~\cite{Prelovsek:2020eiw} & ... & ... & ... & ... & $8\pm5$ & $3929\pm5$ & ... & ... & ... & ... \\
			\multirow{2}{*}{Ref.~\cite{Meng:2020cbk}}
			& ... & ... & ... & ... & ... & ... & $27 \pm 17$ & $4054\pm17$ & $29 \pm 18$ & $4195 \pm 18$ \\
			& ... & ... & ... & ... & ... & ... & $20 \pm 15$ & $4060 \pm 15$ & $22 \pm 16$ & $4202 \pm 16$\\
		\bottomrule[1pt]
		\end{tabular}
\end{table*}

The observations of $Z_{cs}(3985)/Z_{cs}(4000)$ and $Z_{cs}(4120)$ not only enrich the charged charmonium-like states but also make this series of charmonium-like states special. The observations of charmonium-like states near the thresholds of $D^\ast \bar{D}^{(\ast)}$ and $D_s^{(\ast)+}\bar{D}^{\ast}$ motivate theorists to further extend this series to $D_s^{(\ast)} \bar{D}_s^{(\ast)}$ system~\cite{Hidalgo-Duque:2012rqv,Wang:2020htx,Yang:2020nrt,Meng:2020cbk,Prelovsek:2020eiw,Xin:2022bzt,Xie:2022lyw,Qin:2022nof,Giron:2021sla,Lebed:2022vks}. In Table~\ref{tab:be}, we collect the predicted masses of the charmonium-like states near the thresholds of $D_s^{(\ast)+} \bar{D}^{(\ast)}$ and $D_s^{(\ast) +} D_s^{(\ast) -}$, where the $J^{P(C)}$ quantum numbers of $D_s^+ \bar{D}/D_s^+ D_s^-$ systems are $0^{+(+)}$, while the ones of $D_s^{\ast+} \bar{D}^{(\ast )}/D_s^{\ast +} D_s^{(\ast)-}$ system are $1^{+-}$. Very recently, the LHCb Collaboration reported a resonance structure, $X(3960)$, near the threshold of $D_s^+ D_s^-$ threshold in the $B^+\to K^+ D_s^+ D_s^-$ with $J^{PC}=0^{++}$~\cite{LHCb:2022vsv}, which may result from the $D_s^+ D_s^-$ interaction. Considering the components of the molecular states, we use $X^{D_s^{(\ast)}\bar{D}_s^{(\ast)}}$ to represent the possible molecular states composed of $D_s^{(\ast)+}$ and $D_s^{(\ast)-}$ in the present work.

The rich experimental information of this group of near threshold exotic states provides theorists a good opportunity to investigate the deuteron-like hadronic molecular states systematically. To better understand the nature of such kind of charmoniumlike states, the investigations of their production modes are necessary. In our previous work~\cite{Wu:2021cyc}, we have studied the productions of $Z_{cs}(3985)/Z_{cs}(4000)$ in the $B$ and $B_s$ decays by assuming $Z_{cs}(3985)$ and $Z_{cs}(4000)$ are the same $D_s^{+} \bar{D}^{\ast0} + D_s^{\ast +}\bar{D}^0$ molecular state. The estimated branching ratio of $B^+ \to \phi Z_{cs}^+ \to \phi K^+ J/\psi$ is comparable with the measurement from the LHCb Collaboration. In the present work, we assume that the $D^{(*)}_s \bar{D}^{(*)}$ and $D^{(*)}_s \bar{D}^{(*)}_s$  could form bound states, and extend our previous work to study the unobserved charmoniumlike states with strange quark in the $B$ and $B_s$ decays.

The rest of this work is organized as follows. After introduction, we present the model used in the estimations of the $Z_{cs}^{[D_s^{(\ast)} D^{(\ast)}]}$ and $X^{[D_s^{(\ast)}\bar{D}_s^{(\ast)}]}$ productions. The numerical results and discussions are given in Section~\ref{Sec:Num}, and Section~\ref{sec:summary} is devoted to a brief summary.

\begin{figure}[htb]
\begin{tabular}{ccc}
  \centering
  \includegraphics[width=2.8cm]{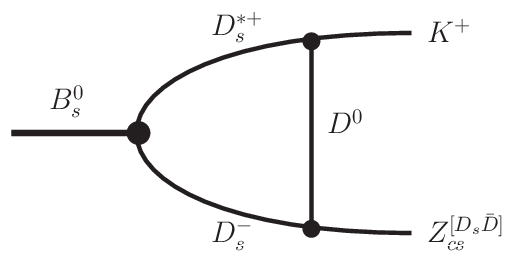}&
  \includegraphics[width=2.8cm]{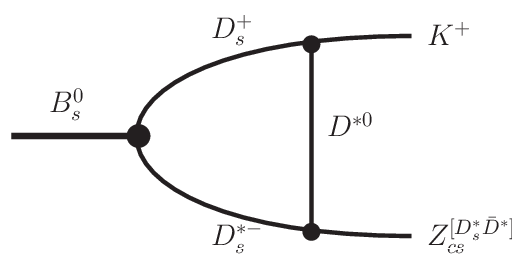}&
  \includegraphics[width=2.8cm]{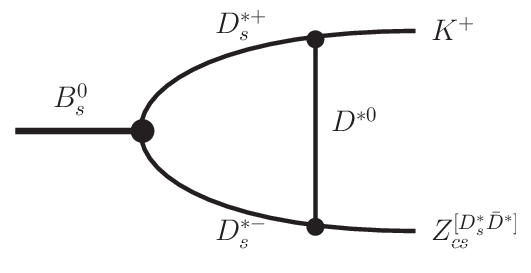}\\
 (a) & (b) & (c) \\
 \end{tabular}
  \caption{Diagrams  contributing to $B_s\rightarrow K + Z_{cs}$. The diagram (a) corresponds to $B_s\rightarrow K + Z_{cs}^{[D_s \bar{D}]}$, the diagrams (b)-(c) correspond to $B_s\rightarrow K + Z_{cs}^{[D^*_s \bar{D}^*]}$.}\label{Fig:TriZcsBs}
\end{figure}

\section{Theoretical framework}
\label{sec:Sec2}

In this work, we systematically investigate the productions of hidden-charm molecular states $\zcs$ and $\xss$ in the $B_s$ and $B$ meson decays. Taking $B_s^0 \to \zcs K^+$ as examples, the initial $B_s^0$ meson couples to a pair of charm-strange mesons, then the pairs of charm-strange mesons transit into the final state $K Z_{cs}^+$ by exchanging a proper charmed meson. The diagrams in Fig.~\ref{Fig:TriZcsBs} reflect this production mechanism in the hadron level. In a similar manner, one can obtain the loop diagrams contributing to $B\rightarrow \phi + Z_{cs}^{[D_s \bar{D}]}$, $B\rightarrow \phi + Z_{cs}^{[D^*_s \bar{D}^*]}$, $B\rightarrow K + X^{[D^{(*)}_s \bar{D}^{(*)}_s]}$, and $B_s\rightarrow \phi + X^{[D^{(*)}_s \bar{D}^{(*)}_s]}$, which are collected in Table~\ref{tab:loops}.

\begin{table}[tb]
	\caption{\label{tab:loops}Loops contributing to $B\rightarrow \phi + Z_{cs}^{[D_s \bar{D}]}$, $B\rightarrow \phi + Z_{cs}^{[D^*_s \bar{D}^*]}$, $B\rightarrow K + X^{[D^{(*)}_s \bar{D}^{(*)}_s]}$ and $B_s\rightarrow \phi + X^{[D^{(*)}_s \bar{D}^{(*)}_s]}$.}
	\renewcommand\arraystretch{1.5}
		\begin{tabular}{p{3cm}<\centering p{5cm}<\centering}
		\toprule[1pt]
		Process & Loops\\
		\midrule[1pt]	
		$B\rightarrow \phi + Z_{cs}^{[D_s \bar{D}]}$  &  $D^+_s \bar{D}^0 D^+_s$ ~, ~ $D^{*+}_s \bar{D}^0 D^+_s$  \\
		$B\rightarrow \phi + Z_{cs}^{[D^*_s \bar{D}^*]}$ &  $D^+_s \bar{D}^{*0} D^{*+}_s$ , $D^{*+}_s \bar{D}^{*0} D^{*+}_s$  \\
		$B\rightarrow K + X^{[D_s \bar{D}_s]}$  &  $\bar{D}^{*0} D^+_s D^{-}_s$     \\
		$B\rightarrow K + X^{[D_s \bar{D}^*_s]}$  & $\bar{D}^{*0} D^{*+}_s D^{-}_s$,  $\bar{D}^{0} D^{+}_s D^{*-}_s$  , $\bar{D}^{*0} D^{+}_s D^{*-}_s$   \\
		$B\rightarrow K + X^{[D^*_s \bar{D}^*_s]}$    & $\bar{D}^{0} D^{*+}_s D^{*-}_s$ , $\bar{D}^{*0} D^{*+}_s D^{*-}_s$  \\
		$B_s\rightarrow \phi + X^{[D_s \bar{D}_s]}$ & $D^+_s D^-_s D^+_s$  ,  $D^{*+}_s D^-_s D^+_s$   \\
		\multirow{2}{*}{$B_s\rightarrow \phi + X^{[D_s \bar{D}^*_s]}$} & $D^+_s D^{*-}_s D^+_s$ , $D^{*+}_s D^{*-}_s D^+_s$ \\
		 &$D^{+}_s D^{-}_s D^{*+}_s$ , $D^{*+}_s D^{-}_s D^{*+}_s$ \\
		$B_s\rightarrow \phi + X^{[D^*_s \bar{D}^*_s]}$  &   $D^{+}_s D^{*-}_s D^{*+}_s$  ,  $D^{*+}_s D^{*-}_s D^{*+}_s$  \\
		\bottomrule[1pt]
		\end{tabular}
	
\end{table}

\subsection{Effective Lagrangian}

The diagrams in Fig.~\ref{Fig:TriZcsBs} and Table~\ref{tab:loops} are evaluated at the hadronic level, where the interactions between hadrons are described by effective Lagrangians. The flavor wave functions of the involved molecular states are,
\begin{equation}
\begin{split}
\left|Z^{[D_s \bar{D}]}_{cs}\right\rangle&=\frac{1}{\sqrt{2}}\left(\left|D^+_s D^-\right\rangle+\left|D^-_s D^+\right\rangle\right)\;,\\
\left|Z^{[D^*_s \bar{D}^*]}_{cs}\right\rangle&=\frac{1}{\sqrt{2}}\left(\left|D^{*+}_s D^{*-}\right\rangle+\left|D^{*-}_s D^{*+}\right\rangle\right)\;,\\
\left|X^{[D_s \bar{D}_s]}\right\rangle&=\left|D^+_s D^{-}_s\right\rangle\;, \\
\left|X^{[D^*_s \bar{D}_s]}\right\rangle&=\frac{1}{\sqrt{2}}\left(\left|D^{*+}_s D^{-}_s\right\rangle+\left|D^{*-}_s D^{+}_s\right\rangle\right)\;, \\
\left|X^{[D^*_s \bar{D}^*_s]}\right\rangle&=\frac{1}{\sqrt{2}}\left(\left|D^{*+}_s D^{*-}_s\right\rangle+\left|D^{*-}_s D^{*+}_s\right\rangle\right)\;.
\end{split}
\end{equation}
The effective couplings of the molecular states to their components are in terms of the following effective Lagrangians,
\begin{equation}\label{eq:lagX}
\begin{split}
 {\cal L}_{Z^{[D_s D]}_{cs}}&=\frac {g_{Z^{[D_s \bar{D}]}_{cs}}} {\sqrt {2}} Z^\dagger_{cs} D_s \bar{D}+h.c.\;, \\
{\cal L}_{Z^{[D^*_s D^*]}_{cs}}&=i\frac {g_{Z^{[D^*_s \bar{D}^*]}_{cs}}} {\sqrt {2}}\varepsilon_{\mu\nu\alpha\beta}\partial^\mu Z^{\dagger\nu}_{cs} D^{*\alpha}_s \bar{D}^{*\beta}+h.c.\;, \\
{\cal L}_{X^{[D_s D_s]}}&=g_{X^{[D_s D_s]}} X^\dagger D_s \bar{D}_s\;,\\
{\cal L}_{X^{[D^*_s D_s]}}&=\frac {g_{X^{[D^*_s D_s]}}} {\sqrt {2}} X^{\dagger\mu} D^{\ast}_{s\mu} \bar{D}_s+h.c. \;,\\
{\cal L}_{X^{[D^*_s D^*_s]}}&=i\frac {g_{X^{[D^*_s D^*_s]}}} {\sqrt {2}}\varepsilon_{\mu\nu\alpha\beta}\partial^\mu X^{\dagger\nu} D^{*\alpha}_s \bar{D}^{*\beta}_s\;,
\end{split}
\end{equation}
where the involved effective coupling constants will be discussed later.

We utilize the naive factorization approach to estimate the decay amplitudes of $B\rightarrow D^{(*)}_s \bar{D}^{(*)}$ and $B_{s}\rightarrow D^{(*)}_s \bar{D}^{(*)}_s$. By applying the effective Hamiltonian at the quark level to the hadron states, we can obtain the parametrized hadronic matrix elements, which are~\cite{Cheng:2003sm,Soni:2021fky},
\begin{equation}\label{Eq:1}
\begin{split}
&\langle0|J_\mu|P(p_1)\rangle= -if_p p_{1\mu}\;,\\
&\langle0|J_\mu|V(p_1,\epsilon)\rangle=f_V \epsilon_\mu m_V \;,\\
&\langle P(p_2)|J_\mu|B_{(s)}(p)\rangle=\Big[P_\mu-\frac{m^2_{B_{(s)}}-m^2_P}{q^2}q_\mu\Big]F_1(q^2)\\
&\qquad+\frac{m^2_{B_{(s)}}-m^2_P}{q^2}q_\mu F_0(q^2) \;,\\
&\langle V(p_2,\epsilon)|J_\mu|B_{(s)}(p)\rangle=\frac{i\epsilon^\nu}{m_{B_{(s)}}+m_V}\Big\{i\varepsilon_{\mu\nu\alpha\beta}P^\alpha q^\beta A_V(q^2)\\
&\qquad+(m_{B_{(s)}}+m_V)^2 g_{\mu\nu}A_1(q^2) -P_\mu P_\nu A_2(q^2)\\
&\qquad-2m_V(m_{B_{(s)}}+m_V)\frac{P_\nu q_\mu}{q^2}[A_3(q^2)-A_0(q^2)]\Big\} \;,
\end{split}
\end{equation}
with $J_\mu=\bar{q}_1 \gamma_\mu(1-\gamma_5)q_2$, $P_\mu=(p+p_2)_\mu$, and $q_\mu=(p-p_2)_\mu$. The form factor $A_3(q^2)$, which is the linear combination of $A_1(q^2)$ and $A_2(q^2)$, is given as~\cite{Cheng:2003sm},
\begin{equation}
A_3(q^2)=\frac{m_{B_{(s)}}+m_V}{2m_V}A_1(q^2)-\frac{m_{B_{(s)}}-m_V}{2m_V}A_2(q^2)\,.
\end{equation}

With Eq.~\eqref{Eq:1}, the amplitudes of $B^0_s\rightarrow D^{(\ast)+}_s D^{(\ast)-}_s$ and $B^+ \rightarrow D^{(\ast)+}_s \bar{D}^{(\ast)0}$ are written as
\begin{equation}\label{Eq:AmpWeak}
\begin{split}
 \mathcal{M}(B^0_s\rightarrow D^{+}_s D^{-}_s)&\equiv\mathcal{A}^{B_s\rightarrow D_s \bar{D}_s}(p_1,p_2)\,, \\
\mathcal{M}(B^0_s\rightarrow D^{+}_s D^{*-}_s)&\equiv\mathcal{A}^{B_s\rightarrow D_s \bar{D}^*_s}_\nu(p_1,p_2)\epsilon^\nu(p_2)\,,\\
\mathcal{M}(B^0_s\rightarrow D^{\ast+}_s D^{-}_s) & \equiv\mathcal{A}^{B_s\rightarrow D^\ast_s \bar{D}_s}_\nu(p_1,p_2)\epsilon^\nu(p_1)\,, \\
\mathcal{M}(B^0_s\rightarrow D^{\ast+}_s D^{\ast-}_s)&\equiv\mathcal{A}^{B_s\rightarrow D^*_s \bar{D}^*_s}_{\mu\nu}(p_1,p_2)\epsilon^\mu(p_1)\epsilon^\nu(p_2)\,,\\
\mathcal{M}(B^+ \rightarrow D^{+}_s \bar{D}^0)&\equiv\mathcal{A}^{B\rightarrow D_s \bar{D}}(p_1,p_2)\,,\\
\mathcal{M}(B^+\rightarrow D^{\ast+}_s \bar{D}^0)&\equiv\mathcal{A}^{B\rightarrow D^\ast_s \bar{D}}_\mu(p_1,p_2)\epsilon^\mu(p_1)\,, \\
\mathcal{M}(B^+\rightarrow D^{+}_s \bar{D}^{*0})&\equiv\mathcal{A}^{B\rightarrow D_s \bar{D}^*}_\nu(p_1,p_2)\epsilon^\nu(p_2)\,, \\
\mathcal{M}(B^+\rightarrow D^{\ast+}_s \bar{D}^{\ast0})&\equiv\mathcal{A}^{B\rightarrow D^*_s \bar{D}^*}_{\mu\nu}(p_1,p_2)\epsilon^\mu(p_1)\epsilon^\nu(p_2)\,.
\end{split}
\end{equation}
Here the expressions of $\mathcal{A}(p_1,p_2)$, $\mathcal{A}_\nu(p_1,p_2)$, and $\mathcal{A}_{\mu\nu}(p_1,p_2)$ are collected in the Appendix~\ref{sec:appendix1}.

For the couplings of the $D^{(*)}_s D^{(*)}K$ and $D^{(*)}_s D^{(*)}_s \phi$, one can construct the Lagrangians relevant to the light vector and pseudoscalar mesons based on the heavy quark limit and chiral symmetry~\cite{Casalbuoni:1996pg,Colangelo:2003sa,Cheng:2004ru}, which are
\begin{equation}\label{eq:LDDV}
\begin{split}
{\cal L} =& -ig_{D^{\ast }D
{\mathcal P}}\left( D^{\dag}_i \partial_\mu {\mathcal P}_{ij} D_j^{\ast \mu}-D_i^{\ast \mu\dagger} \partial_\mu {\mathcal P}_{ij}  D_j\right) \\
& +\frac{1}{2}g_{D^\ast D^\ast {\mathcal P}}\varepsilon _{\mu
\nu \alpha \beta }D_i^{\ast \mu \dag}\partial^\nu {\mathcal P}_{ij}  {\overset{
\leftrightarrow }{\partial }}{\!^{\alpha }} D_j^{\ast \beta } \\
& - ig_{{D}{D}\mathcal{V}} {D}_i^\dagger {\stackrel{\leftrightarrow}{\partial}}{\!_\mu} {D}^j(\mathcal{V}^\mu)^i_j  \\
& -2f_{{D}^*{D}\mathcal{V}} \epsilon_{\mu\nu\alpha\beta}
(\partial^\mu \mathcal{V}^\nu)^i_j
({D}_i^\dagger{\stackrel{\leftrightarrow}{\partial}}{\!^\alpha} {D}^{*\beta j}-{D}_i^{*\beta\dagger}{\stackrel{\leftrightarrow}{\partial}}{\!^\alpha} {D}^j) \\
&+ ig_{{D}^*{D}^*\mathcal{V}} {D}^{*\nu\dagger}_i {\stackrel{\leftrightarrow}{\partial}}{\!_\mu} {D}^{*j}_\nu(\mathcal{V}^\mu)^i_j  \\
& +4if_{{D}^*{D}^*\mathcal{V}} {D}^{*\dagger}_{i\mu}(\partial^\mu \mathcal{V}^\nu-\partial^\nu
\mathcal{V}^\mu)^i_j {D}^{*j}_\nu +{\rm H.c.} \,.
\end{split}
\end{equation}
where the ${D}^{(\ast)}=(D^{(\ast)0},D^{(\ast)+},D^{(\ast)+}_s)$ is the charmed meson triplets, $\mathcal P$ and ${\mathcal V}$ are $3\times 3$ matrix forms of the pseudoscalar and vector mesons, and their concrete forms are,
\begin{eqnarray}
\mathcal{P} &=&
\left(\begin{array}{ccc}
\frac{\pi^{0}}{\sqrt 2}+\alpha\eta+\beta\eta^\prime &\pi^{+} &K^{+}\\
\pi^{-} &-\frac{\pi^{0}}{\sqrt2}+\alpha\eta+\beta\eta^\prime&K^{0}\\
K^{-} &\bar K^{0} &\gamma\eta+\delta\eta^\prime
\end{array}\right),\nonumber\\ \nonumber\\
\mathcal{V} &=& \left(\begin{array}{ccc}\frac{\rho^0} {\sqrt {2}}+\frac {\omega} {\sqrt {2}}&\rho^+ & K^{*+} \\
\rho^- & -\frac {\rho^0} {\sqrt {2}} + \frac {\omega} {\sqrt {2}} & K^{*0} \\
K^{*-}& {\bar K}^{*0} & \phi \\
\end{array}\right) ,
\end{eqnarray}
where the parameters $\alpha$ and $\beta$ related to the mixing angle are defined as
\begin{eqnarray}
\alpha&=&\frac{\cos\theta-\sqrt{2}\sin\theta}{\sqrt{6}},\ \beta=\frac{\sin\theta+\sqrt{2}\cos\theta}{\sqrt{6}},\nonumber\\
\gamma&=&\frac{-2\cos\theta-\sqrt{2}\sin\theta}{\sqrt{6}},\ \delta=\frac{-2\sin\theta+\sqrt{2}\cos\theta}{\sqrt{6}},
\end{eqnarray}
with the mixing angle $\theta=-19.1^\circ$~\cite{MARK-III:1988crp,DM2:1988bfq}.

\subsection{Decay Amplitude}
Using the effective Lagrangians in Eq.~\eqref{eq:LDDV}, the amplitudes for $B^0_s\rightarrow K^+  Z_{cs}$ corresponding to the diagrams in Fig.~\ref{Fig:TriZcsBs} are obtained as,
\begin{equation}\label{eq:amp1}
    \begin{split}
    \mathcal{M}_{a}&= i^3 \int\frac{d^4 q}{(2\pi)^4}\mathcal{A}^{B_s\rightarrow D^*_s \bar{D}_s}_\mu(p_1,p_2)\Big[g_{D^\ast_s D K}p_{3\nu}\Big]\\
&\times \Big[\frac{g_{Z^{[D_s \bar{D}]}_{cs}}}{\sqrt{2}}\Big]\frac{-g^{\mu\nu}+p^{\mu}_1 p^{\nu}_1 /m^2_1}{p_1^2-m^2_1}\frac{1}{p_2^2-m^2_2}\frac{\mathcal{F}(q^2,m_q^2)}{q^2-m^2_q}\, ,\\
\mathcal{M}_{b}&= i^3 \int\frac{d^4 q}{(2\pi)^4}\mathcal{A}^{B_s\rightarrow D_s \bar{D}^*_s}_\mu(p_1,p_2)\Big[-g_{D_s D^* K}p_{3\nu}\Big]\\
&\Big[-\frac{g_{Z^{[D^*_s \bar{D}^*]}_{cs}}}{\sqrt{2}}\varepsilon_{\rho\sigma\alpha\beta}p^\rho_4 \epsilon_{Z_{cs}}^\sigma\Big]\frac{1}{p_1^2-m^2_1}\frac{-g^{\mu\alpha}+p^{\mu}_2 p^{\alpha}_2 /m^2_2}{p_2^2-m^2_2}\\
&\frac{-g^{\nu\beta}+q^{\nu} q^{\beta} /m^2_q}{q^2-m^2_q}\mathcal{F}(q^2,m_q^2)\, ,\\
\mathcal{M}_{c}&= i^3 \int\frac{d^4 q}{(2\pi)^4}\mathcal{A}^{B_s\rightarrow D^*_s \bar{D}^*_s}_{\mu\nu}(p_1,p_2)\Big[-\frac{1}{2}g_{D^\ast_s D^\ast K}\varepsilon_{\rho\tau\kappa\xi}p^\tau_3  \\
&(p_1+q)^\kappa\Big]\Big[-\frac{g_{Z^{[D^*_s \bar{D}^*]}_{cs}}}{\sqrt{2}}\varepsilon_{\omega\sigma\alpha\beta}p^\omega_4 \epsilon_{Z_{cs}}^\sigma\Big]\frac{-g^{\mu\xi}+p^{\mu}_1 p^{\xi}_1 /m^2_1}{p_1^2-m^2_1}\\
&\frac{-g^{\nu\alpha}+p^{\nu}_2 p^{\alpha}_2 /m^2_2}{p_2^2-m^2_2}\frac{-g^{\rho\beta}+q^{\rho} q^{\beta} /m^2_q}{q^2-m^2_q}\mathcal{F}(q^2,m_q^2)\,.
    \end{split}
\end{equation}
The rest of the amplitudes corresponding to the loop diagrams in Table~\ref{tab:loops} can be found in the Appendix~\ref{sec:appendix2}. It should be noted that $\mathcal{M}_{b}$ is proportional to $\epsilon^{\sigma p_3 p_4 q}\varepsilon^{Z_{cs}}_\sigma$, thus it becomes zero after we perform the integral over $q$. Similarly, the corresponding amplitudes for the loop $D^{*+}_s \bar{D}^0 D^+_s$ in $B\rightarrow \phi + Z_{cs}^{[D_s \bar{D}]}$, $\bar{D}^{*0} D^{+}_s D^{*-}_s$ in $B\rightarrow K + X^{[D_s \bar{D}^*_s]}$, $\bar{D}^{0} D^{*+}_s D^{*-}_s$ in $B\rightarrow K + X^{[D^*_s \bar{D}^*_s]}$, and $D^{*+}_s D^-_s D^+_s$ in $B_s\rightarrow \phi + X^{[D_s \bar{D}_s]}$ are also vanish after performing the loop integrals.

In the amplitudes shown in Eq.~\eqref{eq:amp1}, a form factor $\mathcal{F}(q^2,m^2)$ in monopole form is introduced to represent the off-shell effect of the exchanging charmed or charm-strange mesons and to avoid the ultraviolet divergences in the loop integrals. Its concrete form is
\begin{equation}
\mathcal{F}(q^2,m^2) =\frac{m^2 -\Lambda^2}{q^2-\Lambda^2}\,,\label{Eq:A0}
\end{equation}
where $\Lambda=m+\alpha \Lambda_{QCD}$~\cite{Cheng:2004ru} with $\Lambda_{QCD}=220$ MeV. Empirically, the model parameter $\alpha$ should be of the order of unity~\cite{Tornqvist:1993vu, Tornqvist:1993ng,Locher:1993cc,Li:1996yn}, but its accurate value can not be determined by the first principle methods. In practice, we usually check the rationality of the model parameter by comparing our estimations with the corresponding experimental measurements.

With the amplitudes above, the partial width of $B_s\rightarrow K Z_{cs}$ could be estimated by
\begin{equation}
\Gamma_{B_s} = \frac{1}{8\pi} \frac{|\vec{p}|}{m_{B_s}^2}\overline{\big|\mathcal{M}_{B_s\rightarrow K Z_{cs}} \big|^2}\,,
\end{equation}
where the factor $1/(8\pi)$ results from the average of $B_s$ spins and the integration of the phase space, $\vec{p}$ is the momentum of $Z_{cs}$ or $K$ in the rest frame of $B_s$, and $m_{B_s}$ is the mass of $B_s$ meson. The overline indicates the sum over the spins of the final states.

\section{Numerical Results and discussion}
\label{Sec:Num}
\subsection{Coupling constants}

For a shallow bound state, the effective coupling of this state to the two-body channel is related to the probability of finding the two-hadron component in the physical wave function of the bound states, and the effective coupling constant can be determined by~\cite{Weinberg:1965zz,Baru:2003qq}
\begin{equation}
g^2_{\mathrm{eff}}\equiv16\pi(m_1+m_2)^2 \lambda^2\sqrt{\frac{2\Delta E}{\mu}}\,, \label{eq:gzc}
\end{equation}
where $\Delta E=m_1+m_2-M$ denotes the binding energy, $\mu=m_1 m_2/(m_1+m_2)$ is the reduced mass, and $\lambda^2$ gives the probability to find the molecular state in the physical states. In the present molecular scenario, $\lambda=1/ \sqrt{2}$ for $D_s\bar{D}$, $D^*_s\bar{D}^*$, $D^*_s\bar{D}_s$, and $D^*_s\bar{D}^*_s$ system, while $\lambda=1$ for $D_s\bar{D}_s$ sysmtem.

In the non-relativistic limit, the effective Lagrangian of $Z_{cs}^{[D^*_s D^*]}$ and its components should be
\begin{equation}
{\cal L}_{Z^{[D^*_s \bar{D}^*]}_{cs}}=i\frac {g_{Z^{[D^*_s \bar{D}^*]}_{cs}}} {\sqrt {2}} m_{Z_{cs}}\varepsilon_{ijk} Z^{\dagger i}_{cs} D^{*j}_s \bar{D}^{*k}+h.c..
\end{equation}
In this case, the Eq.(\ref{eq:gzc}) should be written as
\begin{equation}
g^2_{Z^{[D^*_s \bar{D}^*]}_{cs}}\equiv16\pi \frac{(m_1+m_2)^2}{m^2_{Z_{cs}}} \lambda^2\sqrt{\frac{2\Delta E}{\mu}}\,. \label{eq:gzcp}
\end{equation}
The couplings of $X^{[D^*_s \bar{D}^*_s]}$ are similar to that of $Z_{cs}^{[D^*_s D^*]}$. We list the coupling constants of $Z^{[D^{(*)}_s \bar{D}^{(*)}]}_{cs}$ and $X^{[D^{(*)}_s \bar{D}^{(*)}_s]}$ with their components in terms of the bingding energy $\Delta E$ in Table~\ref{Table:gzcs}.

\begin{table}
	\caption{\label{Table:gzcs}The coupling constants of $Z^{[D^{(*)}_s \bar{D}^{(*)}]}_{cs}$ and $X^{[D^{(*)}_s \bar{D}^{(*)}_s]}$ with their components in terms of $\Delta E$.}
	\renewcommand\arraystretch{1.5}
		\begin{tabular}{p{2cm}<\centering p{0.95cm}<\centering p{0.95cm}<\centering p{0.95cm}<\centering p{0.95cm}<\centering p{0.95cm}<\centering }
		\toprule[1pt]
			Molecular  & $D_{s}\bar{D}$ & $D^*_{s}\bar{D}^*$ & $D_{s}\bar{D}_s$ & $D_{s}\bar{D}^*_s$ & $D^*_{s}\bar{D}^*_s$ \\
			\midrule[1pt]
			            $\Delta E=10$ MeV  & 7.31 & 1.88 & 10.5 & 7.66 & 1.86  \\
            $\Delta E=30$ MeV  & 9.61 & 2.48 & 13.9 & 10.1 & 2.47  \\
            $\Delta E=50$ MeV  & 10.9 & 2.83 & 15.8 & 11.5 & 2.81 \\
            \bottomrule[1pt]
		\end{tabular}
	\end{table}

In view of the heavy quark limit and chiral symmetry, the coupling constants in Eq.~\eqref{eq:LDDV} satisfy~\cite{Casalbuoni:1996pg,Cheng:2004ru},
\begin{equation}
    \begin{aligned}
        g_{{ D}{ D}V} &= g_{{ D}^*{ D}^*V}=\frac{\beta g_V}{\sqrt{2}} \,, &f_{{ D}^*{ D}V}&=\frac{ f_{{ D}^*{ D}^*V}}{m_{{ D}^*}}=\frac{\lambda g_V}{\sqrt{2}}\,,\\
        g_{{D}^{*} {D} \mathcal{P}}&=\frac{2 g}{f_{\pi}} \sqrt{m_{{D}} m_{{D}^{*}}}\,,  &g_{{D}^{*} {D}^{*} {P}}&=\frac{g_{{ D}^{*} { D} {\mathcal {P}}}}{\sqrt{m_{{D}} m_{{D}^{*}}}}\,,
    \end{aligned}
\end{equation}
where $\beta=0.9$ and $g_V = {m_\rho /f_\pi}$ with $f_\pi = 132$ MeV~\cite{Casalbuoni:1996pg}. The parameters $\lambda = 0.56 \, {\rm GeV}^{-1} $ and $g=0.59$~\cite{Isola:2003fh}, which are estimated by matching the form factor obtained from the light cone sum rule with that calculated from lattice QCD.

As for the form factors in the decay amplitudes of $B\rightarrow D^{(*)}_s \bar{D}^{(*)}$ and $B_{s}\rightarrow D^{(*)}_s \bar{D}^{(*)}_s$, they are usually estimated in the quark model and only known in the space-like regions~\cite{Cheng:2003sm}. One can analytically extend them to the time-like region. In Refs.~\cite{Cheng:2003sm,Soni:2021fky}, the form factor for $B_{(s)} \rightarrow D^{(*)}_{(s)}$ is parameterized as,
\begin{equation}\label{Eq:A1}
F(Q^2)=\frac{F(0)}{1-a\zeta+b\zeta^2}
\end{equation}
with $\zeta=Q^2/m^2_{B_s}$. In Table ~\ref{Tab:PARA1}, the parameters $F(0)$, $a$, and $b$ for $B_{(s)} \rightarrow D^{(*)}_{(s)}$ are collected \footnote{
In Ref.~\cite{Soni:2021fky}, the transition matrix elements of $B_s\rightarrow P/V$ are presented in a different expression
\begin{equation*}
\begin{split}
  \langle P(p_2)|J_\mu|B_{s}(p)\rangle&=F_+(q^2)P^\mu+F_-(q^2)q^\mu\,,\\
\langle V(p_2,\epsilon)|J_\mu|B_{(s)}(p)\rangle&=\frac{\epsilon_\nu}{m+m_2}[-g^{\mu\nu}P\cdot qA_0(q^2)+P^\mu P^\nu A_+(q^2)\\
&+q^\mu P^\nu(q^2)+i\varepsilon^{\mu\nu\alpha\beta}P_\alpha q_\beta V(q^2)]\,,
\end{split}
\end{equation*}
which are identical with the expression in Eq.~\eqref{Eq:1}. By comparing the paramaterizaiton above with that in Eq.~\eqref{Eq:1}, one can find the form factors have the following relation:
\begin{equation}
\begin{split}
F_1(q^2)&=F_+(q^2)\,,\\
F_0(q^2)&=\frac{q^2}{m^2_{B_s}-m^2_P}F_-(q^2)+F_+(q^2)\,,\\
A_V(q^2)&=-iV(q^2),\ A_2(q^2)=iA_+(q^2)\,,\\
A_1(q^2)&=\frac{iP\cdot q}{(m_{B_s}+m_V)^2}A_0(q^2)\,,\\
A_3(q^2)-A_0(q^2)&=\frac{iq^2}{2m_V(m_{B_s}+m_V)}A_-(q^2)\,.
\end{split}
\end{equation}
}.

\begin{table}[tb]
	\caption{\label{Tab:PARA1}The values of the parameters $F(0)$, $a$, and $b$ in the form factors for $B\rightarrow D^{(*)}$~\cite{Cheng:2003sm} and $B_s\rightarrow D^{(*)}_s$~\cite{Soni:2021fky}.}
		\renewcommand\arraystretch{1.5}
		\begin{tabular}{p{0.85cm}<\centering p{0.85cm}<\centering p{0.85cm}<\centering p{0.95cm}<\centering p{0.85cm}<\centering p{0.95cm}<\centering p{0.85cm}<\centering p{0.85cm}<\centering}
		\toprule[1pt]
			$F$ & $F(0)$ & $a$ & $b$ & $F$ & $F(0)$ & $a$ & $b$ \\
			\midrule[1pt]
$F_0$ & 0.67 & 0.65 & 0.00 & $F_1$ & 0.67 & 1.25 & 0.39 \\
$A_V$ & 0.75 & 1.29 & 0.45 & $A_0$ & 0.64 & 1.30 & 0.31 \\
$A_1$ & 0.63 & 0.65 & 0.02 & $A_2$ & 0.61 & 1.14 & 0.52 \\ \midrule[1pt]
$F_+$ & 0.770 & 0.837 & 0.077 & $F_-$ & -0.355 & 0.855 & 0.083 \\
$A_+$ & 0.630 & 0.972 & -0.092 & $A_-$ & -0.756 &  1.001 &  0.116 \\
$A_0$ & 1.564 & 0.442 & -0.178 & $V$ & 0.743 & 1.010 & 0.118 \\
\bottomrule[1pt]
		\end{tabular}
\end{table}

In order to avoid ultraviolet divergence in the loop integrals and evaluate the loop integrals with Feynman parameterization methods, we further parameterize the form factors in the form,
\begin{equation}\label{Eq:A2}
F(Q^2)=F(0)\frac{\Lambda^2_1}{Q^2-\Lambda^2_1}\frac{\Lambda^2_2}{Q^2-\Lambda^2_2}\,,
\end{equation}
where the values of $\Lambda_1$ and $\Lambda_2$ are obtained by fitting Eq.~\eqref{Eq:A1} with Eq.~\eqref{Eq:A2} and the resultant values are list in Table~\ref{Table:PARA2}.

\begin{table}[tb]
	\caption{\label{Table:PARA2}Values of the parameters $\Lambda_1$ and $\Lambda_2$ obtained by fitting Eq.~\eqref{Eq:A1} with Eq.~\eqref{Eq:A2}.}
			\renewcommand\arraystretch{1.5}
		\begin{tabular}{p{1.3cm}<\centering p{1.3cm}<\centering p{0.85cm}<\centering p{0.85cm}<\centering p{0.85cm}<\centering p{0.85cm}<\centering p{0.85cm}<\centering p{0.85cm}<\centering}
		\toprule[1pt]			
		Process& Parameter&   $A_V$ & $A_0$& $A_1$ & $A_2$ &   $F_0$ & $F_1$ \\
			\midrule[1pt]
			\multirow{2}{*}{$B\rightarrow D^{(*)}$} &$\Lambda_1$ & 6.32 & 5.32 & 7.83   & 7.35 &   7.75 & 6.53 \\
			&$\Lambda_2$ & 7.00& 9.41  & 10.99 & 7.35 &  11.00 &  6.84     \\
			\midrule[1pt]
			Process& Parameter&   $A_+$ & $A_-$ & $A_0$ & $V$&  $F_+$ &$F_1$\\ \midrule[1pt]
			\multirow{2}{*}{$B_s\rightarrow D_s^{(*)}$}    &$\Lambda_1$ & 5.48   & 5.77 & 9.75   & 5.74 &   6.30 & 6.25 \\
			 &$\Lambda_2$ & 18.00  & 14.63  & 11.00 & 14.61 &  15.92 &  15.58    \\
			 \bottomrule[1pt]
		\end{tabular}
	\end{table}

\subsection{Branching ratios}

\begin{figure}[t]
  \centering
 \includegraphics[width=8.5cm]{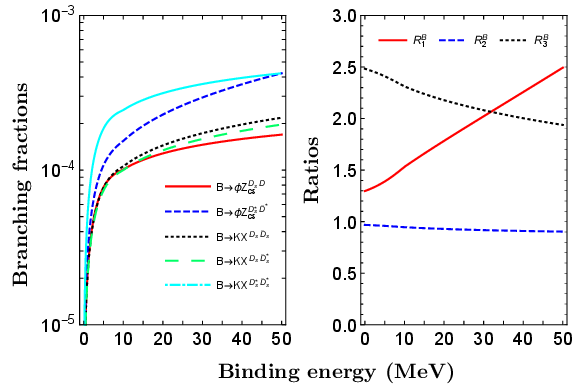}
  \caption{Branching fractions of $B^+\rightarrow \phi \zcs$ and $B^+\rightarrow K^+  \xss$ (left panel) and their ratios (right panel) depending on the binding energy when $\alpha=2$.}\label{Fig:B2}
\end{figure}

\begin{figure}[htb]
  \centering
 \includegraphics[width=8.5cm]{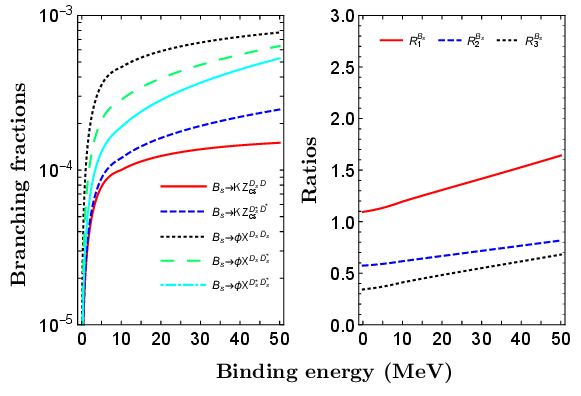}
  \caption{Branching fractions of $B^0_s\rightarrow K^- \zcs$ and $B^0_s\rightarrow \phi \xss$ (left panel) and their ratios (right panel) depending on the binding energy when $\alpha=2$.}\label{Fig:Bs2}
\end{figure}

In our previous work~\cite{Wu:2021cyc}, we take $\alpha=1\sim3$ to estimate the branching fractions of  the processes $B^0_s\rightarrow K^- Z^+_{cs}(3985)$ and $B^+\rightarrow \phi Z^{+}_{cs}(3985)$, and the results are comparable with the measurements from the LHCb Collaboration. In this work, we first take $\alpha=2$ to estimate the productions of $Z_{cs}^{[D_s^{(\ast)} D^{(\ast)}]}$ and $X^{D_s^{(\ast)}\bar{D}_s^{(\ast)}}$ in $B/B_s$ decays.  In the left panels of Figs.~\ref{Fig:B2} and \ref{Fig:Bs2}, we present the branching fractions of $B^+\rightarrow \phi \zcs$, $B^+\rightarrow K^+  \xss$, $B^0_s\rightarrow K^- \zcs$, and $B^0_s\rightarrow \phi  \xss$ in terms of the binding energy when $\alpha=2$. From Eq.~(\ref{eq:gzc}), one can find the effective coupling constant is proportional to $(\Delta E)^{1/4}$, thus, when $\Delta E =  0$, the effective coupling constants $g_{\mathrm{eff}}= 0$, then the branching fraction of $B\to \phi \zcs/K\xss$ and $B_s\to K \zcs/\phi\xss$ are also zero. With the increasing of the binding energy, the branching fractions increase quickly below $\Delta E=5$ MeV and then become weakly dependent on the binding energy. When $\Delta E$ greater than 5 MeV, the magnitudes of branching fractions are of the order of $10^{-4}$, which are at the same order as the branching fractions of $B^0_s\rightarrow K^- Z^+_{cs}(3985)$ and $B^+\rightarrow \phi Z^{+}_{cs}(3985)$ estimated in our previous work~\cite{Wu:2021cyc}. Considering the observations of $Z_{cs}(4000)$ in the decay $B^+ \to J/\psi \phi K^+$, the observations of other hidden charm molecular states in the $B$ and $B_s$ decays should be possible at Belle II and LHCb.

Our estimations indicate that the $\Delta E$ dependences of the relevant branching fractions are very similar. Thus, the ratios of these branching fractions are expected to be weakly dependent on the binding energy. Here, we can categorize the production processes according to two different criterions. Firstly, we can categorize these production processes into four groups by the initial meson and final light meson, and we can define the ratios of the branching fractions for different channels as,
\begin{equation}\label{Eq:ratio1}
\begin{split}
R^B_1&=\frac{\mathcal{B}[B\rightarrow \phi Z^{[D^*_s \bar{D}^*]}_{cs}]}{\mathcal{B}[B\rightarrow\phi Z^{[D_s \bar{D}]}_{cs}]}\,,\\
R^B_2&=\frac{\mathcal{B}[B\rightarrow K X^{[D^*_s \bar{D}_s]}]}{\mathcal{B}[B\rightarrow K X^{[D_s \bar{D}_s]}]}\,,\\
R^B_3&=\frac{\mathcal{B}[B\rightarrow K X^{[D^*_s \bar{D}^*_s]}]}{\mathcal{B}[B\rightarrow K X^{[D_s \bar{D}_s]}]}\,,\\
R^{B_s}_1&=\frac{\mathcal{B}[B_s\rightarrow K Z^{[D^*_s \bar{D}^*]}_{cs}]}{\mathcal{B}[B_s\rightarrow K Z^{[D_s \bar{D}]}_{cs}]}\,,\\
R^{B_s}_2&=\frac{\mathcal{B}[B_s\rightarrow \phi X^{[D^*_s \bar{D}_s]}]}{\mathcal{B}[B_s\rightarrow \phi X^{[D_s \bar{D}_s]}]}\,,\\
R^{B_s}_3&=\frac{\mathcal{B}[B_s\rightarrow \phi X^{[D^*_s \bar{D}^*_s]}]}{\mathcal{B}[B_s\rightarrow \phi X^{[D_s \bar{D}_s]}]}\,.\\
\end{split}
\end{equation}

As shown in the right panels of Figs.~\ref{Fig:B2} and \ref{Fig:Bs2}, one can find that these ratios are very weakly dependent on the binding energy, which are consistent with our expectation. From Fig.~\ref{Fig:B2}, one can find that $R_1^B$ is greater than one, and the maximum of $R_1^B$ is 2.49, which indicate the production rate of $Z_{cs}^{D^\ast_s \bar{D}^\ast}$ is a bit larger than the one of $Z_{cs}^{D_s \bar{D}}$ in the decay $B\to \phi \zcs$. As for $R_2^B$, it is almost independent on the binding energy,  and its value is about 1, which indicates that in the process $B\rightarrow K \xss$, the production fractions of $X^{[D^*_s \bar{D}_s]}$ is comparable to $X^{[D_s \bar{D}_s]}$. Likewise, our estimations also indicate that $R_3^B$ is also greater than one, thus, the production rate of $X^{[D^*_s \bar{D}^*_s]}$ is larger than $X^{[D_s \bar{D}_s]}$ in the process $B\rightarrow K \xss$. In a very similar manner, one can find that the ratio $R_1^{B_s}$ is a bit larger than 1 and its maximum is 1.64, which indicate the production rate of $Z_{cs}^{[D_s^\ast \bar{D}^\ast]}$ are similar to the one of $Z_{cs}^{[D_s \bar{D}]}$ in the process $B_s\to K\zcs$. As for $R_2^{B_s}$ and $R_3^{B_s}$, both of them are smaller than one, moreover, $R_2^{B_s}$ is greater than $R_3^{B_s}$, which indicates the branching fractions satisfy, $\mathcal{B}[B_s \to \phi X^{[D_s^\ast \bar{D}_s^\ast]}]<\mathcal{B}[B_s \to \phi X^{[D_s^\ast \bar{D}_s]}]<\mathcal{B}[B_s \to \phi X^{[D_s \bar{D}_s]}]$.

\begin{figure}[t]
  \centering
 \includegraphics[width=8cm]{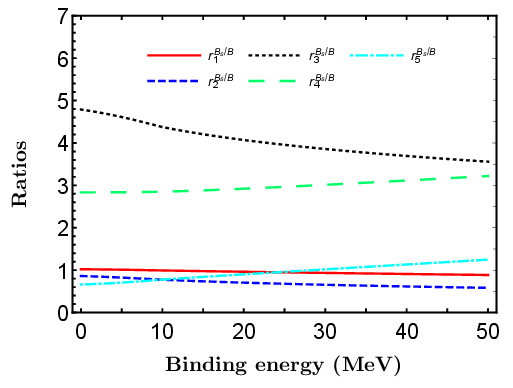}
  \caption{The ratios defined in Eq.~(\ref{Eq:ratio2}) when $\alpha=2$.}\label{Fig:ratio2}
\end{figure}

We can also categorize the discussed processes into five groups by the involved hidden-charm molecular states, and the relevant ratios are defined as
\begin{equation}\label{Eq:ratio2}
\begin{split}
r^{B_s/B}_1&=\frac{\mathcal{B}[B_s\rightarrow K Z^{[D_s \bar{D}]}_{cs}]}{\mathcal{B}[B\rightarrow\phi Z^{[D_s \bar{D}]}_{cs}]}\,,\\
r^{B_s/B}_2&=\frac{\mathcal{B}[B_s\rightarrow K Z^{[D^*_s \bar{D}^*]}_{cs}]}{\mathcal{B}[B\rightarrow\phi Z^{[D^*_s \bar{D}^*]}_{cs}]}\,,\\
r^{B_s/B}_3&=\frac{\mathcal{B}[B_s\rightarrow \phi X^{[D_s \bar{D}_s]}]}{\mathcal{B}[B\rightarrow K X^{[D_s \bar{D}_s]}]}\,,\\
r^{B_s/B}_4&=\frac{\mathcal{B}[B_s\rightarrow \phi X^{[D^*_s \bar{D}_s]}]}{\mathcal{B}[B\rightarrow K X^{[D^*_s \bar{D}_s]}]}\,,\\
r^{B_s/B}_5&=\frac{\mathcal{B}[B_s\rightarrow \phi X^{[D^*_s \bar{D}^*_s]}]}{\mathcal{B}[B\rightarrow K X^{[D^*_s \bar{D}^*_s]}]}\,.\\
\end{split}
\end{equation}

In Fig.~\ref{Fig:ratio2}, we plot the second type of the ratios defined in Eq.~\eqref{Eq:ratio2} depending on the binding energy with $\alpha=2$. Similar to the first type of ratios shown in the right panels of Figs.~\ref{Fig:B2} and \ref{Fig:Bs2}, the second ratios are also very weakly dependent on the binding energy. In particular, the ratio $r_1^{B_s/B}$, $r_2^{B_s/B}$ and $r_5^{B_s/B}$ are close to one, which indicate the production rates of $Z_{cs}^{D_s\bar{D}}$,  $Z_{cs}^{D_s^\ast\bar{D}^\ast}$, and $X^{D_s^\ast \bar{D}_s^\ast}$ in $B$ and $B_s$ decays are very similar.  $r_3^{B_s/B}$ and $r_4^{B_s/B}$ are greater than one, which means the production rates of $X^{D_s \bar{D}_s}$ and $X^{D_s^\ast \bar{D}_s}$ in the $B_s$ decay is larger than the one in the $B$ decay, and these hidden-charm molecular states may be more potentially be observed in the $B_s$ decay processes.

\begin{table*}[t]
	\caption{\label{tab:br}Predicted branching ratios (in units of $10^{-4}$) when we take $\alpha=1\,,2\,,3$ and $\Delta E=10\,,30\,,50$ MeV.}
	\renewcommand\arraystretch{1.5}
		\begin{tabular}{p{1.3cm}<\centering p{0.95cm}<\centering p{0.95cm}<\centering p{0.95cm}<\centering p{0.95cm}<\centering p{0.95cm}<\centering p{0.95cm}<\centering p{0.95cm}<\centering p{0.95cm}<\centering p{0.95cm}<\centering p{0.95cm}<\centering p{0.95cm}<\centering p{0.95cm}<\centering p{0.95cm}<\centering p{0.95cm}<\centering p{0.95cm}<\centering}
		\toprule[1pt]		
			Process & \multicolumn{3}{c}{$B\rightarrow \phi Z^{[D_s \bar{D}]}_{cs}$} & \multicolumn{3}{c}{$B\rightarrow\phi Z^{[D^*_s \bar{D}^*]}_{cs}$} & \multicolumn{3}{c}{$B\rightarrow K X^{[D_s \bar{D}_s]}$} & \multicolumn{3}{c}{$B\rightarrow K X^{[D^*_s \bar{D}_s]}$} & \multicolumn{3}{c}{$B\rightarrow K X^{[D^*_s \bar{D}^*_s]}$} \\
			\midrule[1pt]
			Parameter  & $\alpha=1$ & $\alpha=2$ & $\alpha=3$ & $\alpha=1$ & $\alpha=2$ & $\alpha=3$ & $\alpha=1$ & $\alpha=2$ & $\alpha=3$ & $\alpha=1$ & $\alpha=2$ & $\alpha=3$ & $\alpha=1$ & $\alpha=2$ & $\alpha=3$\\
			$\Delta E=10$ & $0.50$ & $1.01$ & $1.44$ & $0.55$ & $1.54$ & $2.89$ & $0.33$ & $1.06$ & $2.12$ & $0.41$ &$1.00$ &$1.64$ &$1.21$ &$2.44$ &$3.50$\\
			$\Delta E=30$ & $0.67$ & $1.46$ & $2.17$ & $0.98$ & $2.95$ & $5.68$ & $0.51$ & $1.73$ & $3.55$ & $0.61$ &$1.59$ &$2.68$ &$1.63$ &$3.60$ &$5.37$\\
			$\Delta E=50$ & $0.73$ & $1.70$ & $2.59$ & $1.34$ & $4.23$ & $8.27$ & $0.62$ & $2.18$ & $4.53$ & $0.73$ &$1.97$ &$3.38$ &$1.82$ &$4.23$ &$6.47$\\
			\midrule[1pt]
			Process & \multicolumn{3}{c}{$B_s\rightarrow K Z^{[D_s \bar{D}]}_{cs}$} & \multicolumn{3}{c}{$B_s\rightarrow K Z^{[D^*_s \bar{D}^*]}_{cs}$} & \multicolumn{3}{c}{$B_s\rightarrow \phi X^{[D_s \bar{D}_s]}$} & \multicolumn{3}{c}{$B_s\rightarrow \phi X^{[D^*_s \bar{D}_s]}$} & \multicolumn{3}{c}{$B_s\rightarrow \phi X^{[D^*_s \bar{D}^*_s]}$} \\
			\midrule[1pt]
			Parameter  & $\alpha=1$ & $\alpha=2$ & $\alpha=3$ & $\alpha=1$ & $\alpha=2$ & $\alpha=3$ & $\alpha=1$ & $\alpha=2$ & $\alpha=3$ & $\alpha=1$ & $\alpha=2$ & $\alpha=3$ & $\alpha=1$ & $\alpha=2$ & $\alpha=3$ \\
			$\Delta E=10$ & $0.50$ & $1.00$ & $1.43$ & $0.48$ & $1.20$ & $2.12$ & $2.31$ & $4.62$ & $6.60$ & $1.34$ &$2.85$ &$5.21$ &$0.61$ &$1.90$ &$3.73$\\
			$\Delta E=30$ & $0.62$ & $1.36$ & $2.03$ & $0.72$ & $1.93$ & $3.56$ & $3.06$ & $6.68$ & $9.91$ & $1.83$ &$4.79$ &$9.57$ &$1.13$ &$3.67$ &$7.32$\\
			$\Delta E=50$ & $0.65$ & $1.50$ & $2.30$ & $0.87$ & $2.47$ & $4.65$ & $3.37$ & $7.76$ & $11.8$ & $2.15$ &$6.35$ &$13.3$ &$1.58$ &$5.28$ &$10.6$\\
			\bottomrule[1pt]
		\end{tabular}
	\end{table*}

As discussed in Ref.~\cite{Wu:2021cyc}, when we take the model parameter $\alpha=1\sim 3$, the estimated branching fraction of $B^+ \to J/\psi \phi K^+$ is comparable with the experimental measurement. In the above estimations, we mainly focus on the binding energy dependences of the branching fractions, and in the following, we further discuss the model parameter dependence of the branching fractions. In Table~\ref{tab:br}, we list the predicted branching ratios (in units of $10^{-4}$) with $\alpha=1\,,2\,,3$ and $\Delta E=10\,,30\,,50$ MeV, respectively. From the Table, one can find that the branching fractions are also weakly dependent on the model parameter $\alpha$ when we fix the binding energy, and most of the estimated branching fractions of the considered processes are of the order of $10^{-4}$. As shown in Figs.~\ref{Fig:B2} and \ref{Fig:Bs2}, the binding energy dependences of the branching fractions are rather weak when $\Delta_E>5 $ MeV, thus we can fix $\Delta_E$ to discuss the $\alpha$ dependences of the branching fractions. For example, when we take $\Delta E=30$ MeV, we can obtain the following ratios,
\begin{eqnarray}
 &&\mathcal{B}[B\rightarrow K X^{[D_s \bar{D}_s]}]:\mathcal{B}[B\rightarrow K X^{[D^*_s \bar{D}_s]}]:\mathcal{B}[B\rightarrow K X^{[D^*_s \bar{D}^*_s]}]\nonumber \\
&&=1.09^{+1.37}_{-1.02}:1:2.26^{+1.91}_{-1.87}\nonumber, \\
&&\mathcal{B}[B_s\rightarrow \phi X^{[D_s \bar{D}_s]}]:\mathcal{B}[B_s\rightarrow \phi X^{[D^*_s \bar{D}_s]}]:\mathcal{B}[B_s\rightarrow \phi X^{[D^*_s \bar{D}^*_s]}] \nonumber\\
&&=1.39^{+1.55}_{-1.15}:1:0.77^{+1.08}_{-0.71}\,,
\end{eqnarray}
where the center value is estimated by taking $\alpha=2$ and the uncertainties result from variation of model parameter $\alpha$ from 1 to 3.
Moreover, combining with the results in Ref.~\cite{Wu:2021cyc}, one can conclude,
\begin{eqnarray}
&&\mathcal{B}[B\rightarrow \phi Z^{[D_s \bar{D}]}_{cs}]:\mathcal{B}[B\rightarrow \phi Z^{[D^*_s \bar{D}]}_{cs}]:\mathcal{B}[B\rightarrow \phi Z^{[D^*_s \bar{D}^*]}_{cs}] \nonumber \\
&&=3.11^{+1.88}_{-1.92}:1:6.28^{+6.24}_{-4.59}\nonumber, \\
&&\mathcal{B}[B_s\rightarrow K Z^{[D_s \bar{D}]}_{cs}]:\mathcal{B}[B_s\rightarrow K Z^{[D^*_s \bar{D}]}_{cs}]:\mathcal{B}[B_s\rightarrow K Z^{[D^*_s \bar{D}^*]}_{cs}]\nonumber \\
&&=0.66^{+0.36}_{-0.44}:1:0.93^{+0.82}_{-0.69}\,.
\end{eqnarray}
All the above relations could be tested by future experiments at Belle II and LHCb, which can help us to better understand the hidden-charm molecular states with strange quark.

\section{Summary}
\label{sec:summary}

In the present work, we perform a systematical investigation of the production of hidden-charm tetraquarks with strange quark in the $B_s$ and $B$ meson decays. The present estimations indicate that the branching fractions of the relevant processes are of the order of $10^{-4}$, which are at the same order of the branching ratios of $B^0_s\rightarrow K^- Z^+_{cs}(3985)$ and $B^+\rightarrow \phi Z^{+}_{cs}(3985)$ estimated in our previous work~\cite{Wu:2021cyc}. Considering the observation of $Z_{cs}(3985)$ in the process $B^+ \to J/\psi \phi K^+$, one can conclude that the observations of the discussed hidden charm molecular states with strange quark in the $B$ and $B_s$ decays should be possible at Belle II and LHCb.

In addition, our estimations indicate the binding energy $\Delta E$ and model parameter $\alpha$ dependences of the branching fractions are very similar, thus, one can expect the ratios of the branching fractions should be weakly dependent on the model parameter and binding energy. According to the values of the relative ratios, we propose the promising channels of searching these hidden charm molecular states with strange quark, for example, the process $B_s\rightarrow \phi X^{[D_s \bar{D}_s]}$ should be more suitable for observing the $X^{[D_s \bar{D}_s]}$ than the process $B \rightarrow K X^{[D_s \bar{D}_s]}$. Moreover, we also predict the ratios of the branching fractions of $B\rightarrow KX$ and $B_s\rightarrow \phi X$ for each $\xss$, and the ratios of the branching fractions of  $B\rightarrow \phi Z_{cs}$ and $B_s\rightarrow K Z_{cs}$ for each $\zcs$,  which could be tested by future experimental measurements at Belle II and LHCb.

\begin{acknowledgments}
Q. W is grateful to Professor Shi-Lin Zhu for very helpful discussions. This work is supported by the National Natural Science Foundation of China (NSFC) under Grant No.11775050, 12175037, 12105153, 11835015, and 12075133 and the Natural Science Foundation of Shandong province under the Grant No. ZR2021MA082, and ZR2022ZD26. It is also partly supported by Taishan Scholar Project of Shandong Province (Grant No. tsqn202103062), the Higher Educational Youth Innovation Science and Technology Program Shandong Province (Grant No. 2020KJJ004).
\end{acknowledgments}

\onecolumngrid
	\appendix
	\section{The expressions of $\mathcal{A}(p_1,p_2)$\,, $\mathcal{A}_\nu(p_1,p_2)$\,, and $\mathcal{A}_{\mu\nu}(p_1,p_2)$}\label{sec:appendix1}
Here we collect all the functions used in Eq. ~\eqref{Eq:AmpWeak}, which are,
	\begin{equation*}
		\begin{split}
			\mathcal{A}^{B_s\rightarrow D_s \bar{D}_s}(p_1,p_2)&=-\frac{iG_F}{\sqrt{2}}V_{cb}V^\ast_{cs}a_1 f_{D_s}(m^2_{B_s}-m^2_{D_s})F_0^{B_s D_s}(p^2_1) \, ,\\
			\mathcal{A}^{B_s\rightarrow D^*_s \bar{D}_s}(p_1,p_2)&=\frac{2G_F}{\sqrt{2}}V_{cb}V^\ast_{cs}a_1 f_{D^*_s}m_{D^\ast_s} p_{2\mu} F_1^{B_s D_s}(p^2_1)  \, , \\
			\mathcal{A}^{B_s\rightarrow D_s \bar{D}^*_s}_\nu(p_1,p_2)&=\frac{G_F}{\sqrt{2}}V_{cb}V^\ast_{cs}a_1 f_{D_s}\frac{1}{m_{B_s}+m_{D^\ast_s}} \Big\{(m_{B_s}+m_{D^\ast_s})^2 g_{\mu\nu}p^\mu_{1}A_1^{B_s D^\ast_s}(p^2_1)-(p_1+2p_2)_\mu (p_1+2p_2)_\nu p^\mu_{1} A_2^{B_s D^\ast_s}(p^2_1)\\
			&-2m_{D^\ast_s}(m_{B_s}+m_{D^\ast_s})(p_1+2p_2)_\nu [A_3^{B_s D^\ast_s}(p^2_1)-A_0^{B_s D^\ast_s}(p^2_1)]\Big\}  \, , \\
			\mathcal{A}^{B_s\rightarrow D^*_s \bar{D}^*_s}_{\mu\nu}(p_1,p_2)&=\frac{G_F}{\sqrt{2}}V_{cb}V^\ast_{cs}a_1 f_{D^*_s}m_{D^\ast_s}\frac{i}{m_{B_s}+m_{D^\ast_s}} \Big\{i\varepsilon_{\mu\nu\alpha\beta}(p_1+2p_2)^\alpha p^\beta_1 A_V^{B_s D^\ast_s}(p^2_1)+(m_{B_s}+m_{D^\ast_s})^2 g_{\mu\nu}A_1^{B_s D^\ast_s}(p^2_1) \\
			&-(p_1+2p_2)_\mu (p_1+2p_2)_\nu A_2^{B_s D^\ast_s}(p^2_1)\Big\}  \, ,\\
			\mathcal{A}^{B\rightarrow D_s \bar{D}}(p_1,p_2)&=-\frac{iG_F}{\sqrt{2}}V_{cb}V^\ast_{cs}a_1 f_{D_s}(m^2_{B}-m^2_{D})F_0^{B D}(p^2_1)  \, , \\
			\mathcal{A}_\mu^{B\rightarrow D^*_s \bar{D}}(p_1,p_2)&=\frac{2G_F}{\sqrt{2}}V_{cb}V^\ast_{cs}a_1 f_{D^*_s}m_{D^\ast_s} p_{2\mu} F_1^{B D}(p^2_1)  \, , \\
			\mathcal{A}^{B\rightarrow D_s \bar{D}^*}_\nu(p_1,p_2)&=\frac{G_F}{\sqrt{2}}V_{cb}V^\ast_{cs}a_1 f_{D_s}\frac{1}{m_{B}+m_{D^\ast}} \Big\{(m_{B}+m_{D^\ast})^2 g_{\mu\nu}p^\mu_{1}A_1^{B D^\ast}(p^2_1) -(p_1+2p_2)_\mu (p_1+2p_2)_\nu p^\mu_{1} A_2^{B D^\ast}(p^2_1) \\
			&-2m_{D^\ast}(m_{B}+m_{D^\ast})(p_1+2p_2)_\nu [A_3^{B D^\ast}(p^2_1)-A_0^{B D^\ast}(p^2_1)]\Big\}  \, , \\
			\mathcal{A}^{B\rightarrow D^*_s \bar{D}^*}_{\mu\nu}(p_1,p_2)&=i\frac{G_F}{\sqrt{2}}V_{cb}V^\ast_{cs}a_1 f_{D^*_s}\frac{m_{D^\ast_s}}{m_{B_s}+m_{D^\ast}} \Big\{i\varepsilon_{\mu\nu\alpha\beta}(p_1+2p_2)^\alpha p^\beta_1 A_V^{B D^\ast}(p^2_1) +(m_{B}+m_{D^\ast})^2 g_{\mu\nu}A_1^{B D^\ast}(p^2_1) \\
			&-(p_1+2p_2)_\mu (p_1+2p_2)_\nu A_2^{B D^\ast}(p^2_1)\Big\} \, .
		\end{split}
	\end{equation*}

	\section{Decay amplitude}\label{sec:appendix2}
	The amplitudes corresponding to the loop diagrams contributing to process $B \rightarrow \phi + Z_{cs}$ are,
		\begin{equation*}
		\begin{split}
			\mathcal{M}_{D^+_s \bar{D}^0 D^+_s}&=i^3 \int\frac{d^4 q}{(2\pi)^4}\mathcal{A}^{B\rightarrow D_s \bar{D}}(p_1,p_2)\Big[-g_{D_s D_s \phi}(p_{1\rho}+q_\rho)\epsilon^\rho_{\phi}\Big] \Big[\frac{g_{Z^{[D_s \bar{D}]}_{cs}}}{\sqrt{2}}\Big] \\ & \times \frac{1}{p^2_1-m^2_1}\frac{1}{p^2_2-m^2_2}\frac{1}{q^2-m^2_q}\mathcal{F}(q^2,m_q^2) \, ,\\
			\mathcal{M}_{D^{*+}_s \bar{D}^0 D^+_s}&=i^3 \int\frac{d^4 q}{(2\pi)^4}\mathcal{A}_{\mu}^{B\rightarrow D^*_s \bar{D}}(p_1,p_2)\Big[-2f_{D_s D^*_s \phi}\varepsilon_{\rho\tau\delta\xi}p^\rho_3 \epsilon^\tau_{\phi}(p^\delta_1+q^\delta)\Big]\Big[\frac{g_{Z^{[D_s \bar{D}]}_{cs}}}{\sqrt{2}}\Big]\\
			&\times\frac{-g^{\mu\xi}+p^{\mu}_1 p^{\xi}_1 /m^2_1}{p^2_1-m^2_1} \frac{1}{p^2_2-m^2_2}\frac{1}{q^2-m^2_q}\mathcal{F}(q^2,m_q^2) \, ,\\
			\mathcal{M}_{D^+_s \bar{D}^{*0} D^{*+}_s}&=i^3 \int\frac{d^4 q}{(2\pi)^4}\mathcal{A}_\mu^{B\rightarrow D_s \bar{D}^*}(p_1,p_2)\Big[2f_{D_s D^*_s \phi}\varepsilon_{\rho\tau\kappa\xi} p^\rho_3 \epsilon^\tau_{\phi}(p_1+q)^\kappa\Big]\Big[-\frac{g_{Z^{[D^*_s \bar{D}^*]}_{cs}}}{\sqrt{2}}\varepsilon_{\omega\sigma\alpha\beta}p^\omega_4 \epsilon_{Z_{cs}}^\sigma\Big] \\
			&\times  \frac{1}{p^2_1-m^2_1}\frac{-g^{\mu\beta}+p^{\mu}_2 p^{\beta}_2 /m^2_2}{p^2_2-m^2_2}\frac{-g^{\xi\alpha}+q^{\xi} q^{\alpha} /m^2_q}{q^2-m^2_q}  \mathcal{F}(q^2,m_q^2) \, ,\\
			\mathcal{M}_{D^{*+}_s \bar{D}^{*0} D^{*+}_s}&=i^3 \int\frac{d^4 q}{(2\pi)^4}\mathcal{A}_{\mu\nu}^{B\rightarrow D^*_s \bar{D}^*}(p_1,p_2)\Big[g_{D^\ast_s D^\ast_s \phi}g_{\tau\theta}(p_{1\kappa}+q_{\kappa})\epsilon^\kappa_{\phi}+4f_{D^\ast_s D^\ast_s \phi}g_{\tau\kappa}g_{\theta\nu}(p^\nu_3 \epsilon^\kappa_{\phi}-p^\kappa_3 \epsilon^\nu_{\phi})\Big]\nonumber \\
			&\times\Big[-\frac{g_{Z^{[D^*_s \bar{D}^*]}_{cs}}}{\sqrt{2}}\varepsilon_{\omega\sigma\alpha\beta}p^\omega_4 \epsilon_{Z_{cs}}^\sigma\Big]\frac{-g^{\mu\tau}+p^{\mu}_1 p^{\tau}_1 /m^2_1}{p^2_1-m^2_1} \frac{-g^{\nu\beta}+p^{\nu}_2 p^{\beta}_2 /m^2_2}{p^2_2-m^2_2}\frac{-g^{\theta\alpha}+q^{\theta} q^{\alpha} /m^2_q}{q^2-m^2_q}\mathcal{F}(q^2,m_q^2) \, .
		\end{split}
	\end{equation*}

The amplitudes corresponding to the loop diagrams contributing to process $B \rightarrow K + X$ are
	\begin{equation*}
		\begin{split}
			\mathcal{M}_{\bar{D}^{*0} D^+_s D^{-}_s}&=i^3 \int\frac{d^4 q}{(2\pi)^4}\mathcal{A}_\mu^{B\rightarrow \bar{D}^* D_s}(p_1,p_2)\Big[-g_{D^* D_s K}p_{3\rho}\Big] \Big[g_{X^{[D_s D_s]}}\Big]\\
			&\times\frac{-g^{\mu\rho}+p^{\mu}_1 p^{\rho}_1 /m^2_1}{p^2_1-m^2_1}\frac{1}{p^2_2-m^2_2}\frac{1}{q^2-m^2_q} \mathcal{F}(q^2,m_q^2)\, ,\\
			\mathcal{M}_{\bar{D}^{*0} D^{*+}_s D^{-}_s}&=i^3 \int\frac{d^4 q}{(2\pi)^4}\mathcal{A}^{B\rightarrow \bar{D}^* D^*_s}_{\mu\nu}(p_1,p_2)\Big[-g_{D^* D_s K}ip_{3\rho}\Big] \Big[\frac{ g_{X^{[D^*_s D_s]}}}{\sqrt{2}}\epsilon^{Z_{cs}}_\sigma\Big]\\
			&\times \frac{-g^{\mu\rho}+p^{\mu}_1 p^{\rho}_1 /m^2_1}{p_1^2-m^2_1}\frac{-g^{\nu\sigma}+p^{\nu}_2 p^{\sigma}_2 /m^2_2}{p^2_2-m^2_2} \frac{1}{q^2-m^2_q}\mathcal{F}(q^2,m_q^2) \, ,\\
			\mathcal{M}_{\bar{D}^{0} D^{+}_s D^{*-}_s}&=i^3 \int\frac{d^4 q}{(2\pi)^4}\mathcal{A}^{B\rightarrow \bar{D}^* D^*_s}(p_1,p_2)\Big[g_{D D^*_s K}p_{3\rho}\Big]  \Big[\frac{ g_{X^{[D^*_s D_s]}}}{\sqrt{2}}\epsilon^{Z_{cs}}_\sigma\Big]\\
			&\times \frac{1}{p_1^2-m^2_1}\frac{1}{p^2_2-m^2_2}\frac{-g^{\rho\sigma}+q^{\rho} q^{\sigma} /m^2_q}{q^2-m^2_q} \mathcal{F}(q^2,m_q^2)\, , \\
			\mathcal{M}_{\bar{D}^{*0} D^{+}_s D^{*-}_s}&=i^3 \int\frac{d^4 q}{(2\pi)^4}\mathcal{A}^{B\rightarrow \bar{D}^* D_s}_{\mu}(p_1,p_2)\Big[-\frac{1}{2}g_{D^\ast_s D^\ast K}\varepsilon_{\rho\tau\kappa\xi}p^\tau_3 (q+p_1)^\kappa\Big]\Big[\frac{g_{X^{[D^*_s D_s]}}}{\sqrt{2}}\epsilon^{Z_{cs}}_\sigma\Big]\\
			&\times\frac{-g^{\mu\rho}+p^{\mu}_1 p^{\rho}_1 /m^2_1}{p_1^2-m^2_1} \frac{1}{p_2^2-m^2_2}\frac{-g^{\xi\sigma}+q^{\xi} q^{\sigma} /m^2_q}{q^2-m^2_q}\mathcal{F}(q^2,m_q^2) \, ,\\
			\mathcal{M}_{\bar{D}^{0} D^{*+}_s D^{*-}_s}&=i^3 \int\frac{d^4 q}{(2\pi)^4}\mathcal{A}^{B\rightarrow \bar{D} D^*_s}_\nu(p_1,p_2)\Big[g_{D D^*_s K}p_{3\rho}\Big] \Big[-\frac{g_{X^{[D^*_s D^*_s]}}}{\sqrt{2}}\varepsilon_{\omega\sigma\lambda\xi}p^\omega_4 \epsilon_{Z_{cs}}^\sigma\Big]\\
			&\times \frac{1}{p_1^2-m^2_1}\frac{-g^{\nu\lambda}+p^{\nu}_2 p^{\lambda}_2 /m^2_2}{p^2_2-m^2_2} \frac{-g^{\rho\xi}+q^{\rho} q^{\xi} /m^2_q}{q^2-m^2_q}\mathcal{F}(q^2,m_q^2) \, ,\\
			\mathcal{M}_{\bar{D}^{*0} D^{*+}_s D^{*-}_s}&=i^3 \int\frac{d^4 q}{(2\pi)^4}\mathcal{A}^{B\rightarrow \bar{D}^* D^*_s}_{\mu\nu}(p_1,p_2)\Big[-\frac{1}{2}g_{D^\ast_s D^\ast K}\varepsilon_{\rho\tau\kappa\xi}p^\tau_3 (q+p_1)^\kappa\Big]\Big[-\frac{g_{X^{[D^*_s D^*_s]}}}{\sqrt{2}}\varepsilon_{\omega\sigma\lambda\delta}p^\omega_4 \epsilon_{Z_{cs}}^\sigma\Big] \\
			&\times\frac{-g^{\mu\rho}+p^{\mu}_1 p^{\rho}_1 /m^2_1}{p_1^2-m^2_1}\frac{-g^{\nu\lambda}+p^{\nu}_2 p^{\lambda}_2 /m^2_2}{p_2^2-m^2_2} \frac{-g^{\xi\delta}+q^{\xi} q^{\delta} /m^2_q}{q^2-m^2_q}\mathcal{F}(q^2,m_q^2) \, .
		\end{split}
	\end{equation*}

	The amplitudes corresponding to the loop diagrams contributing to process $B^0_s \rightarrow \phi + X$ are
	\begin{equation*}
		\begin{split}
			\mathcal{M}_{D^+_s D^-_s D^+_s}&=i^3 \int\frac{d^4 q}{(2\pi)^4}\mathcal{A}_\nu^{B_s\rightarrow D_s \bar{D}_s}(p_1,p_2) \Big[-g_{D_s D_s \phi}(p_{1\rho}+q_\rho)\epsilon^\rho_{\phi}\Big]\Big[g_{X^{[D_s D_s]}}\Big]\\
			&\times\frac{1}{p^2_1-m^2_1}\frac{1}{p^2_2-m^2_2}\frac{1}{q^2-m^2_q}\mathcal{F}(q^2,m_q^2) \, ,\\
			\mathcal{M}_{D^{*+}_s D^-_s D^+_s}&=i^3 \int\frac{d^4 q}{(2\pi)^4}\mathcal{A}_{\mu}^{B_s\rightarrow D^*_s \bar{D}_s}(p_1,p_2)\Big[-2f_{D_s D^*_s \phi}\varepsilon_{\rho\tau\delta\xi}p^\rho_3 \epsilon^\tau_{\phi}(p^\delta_1+q^\delta)\Big]\Big[g_{X^{[D_s D_s]}}\Big]\\
			&\times \frac{-g^{\mu\xi}+p^{\mu}_1 p^{\xi}_1 /m^2_1}{p^2_1-m^2_1} \frac{1}{p^2_2-m^2_2}\frac{1}{q^2-m^2_q}\mathcal{F}(q^2,m_q^2) \, ,\\
			\mathcal{M}_{D^+_s D^{*-}_s D^+_s}&=i^3 \int\frac{d^4 q}{(2\pi)^4}\mathcal{A}_\mu^{B_s\rightarrow D_s \bar{D}^*_s}(p_1,p_2)\Big[-g_{D_s D_s \phi} \times(p_{1\rho}+q_\rho)\epsilon^\rho_{\phi}\Big]\Big[\frac{g_{X^{[D^*_s D_s]}}}{\sqrt{2}}\epsilon^{Z_{cs}}_\nu\Big]\\
			&\times \frac{1}{p^2_1-m^2_1} \frac{-g^{\mu\nu}+p^{\mu}_2 p^{\nu}_2 /m^2_2}{p^2_2-m^2_2}\frac{1}{q^2-m^2_q}\mathcal{F}(q^2,m_q^2) \, ,\\
			\mathcal{M}_{D^{*+}_s D^{*-}_s D^+_s}&=i^3 \int\frac{d^4 q}{(2\pi)^4}\mathcal{A}_{\mu\nu}^{B_s\rightarrow D^*_s \bar{D}^*_s}(p_1,p_2)\Big[-2f_{D_s D^*_s \phi}\varepsilon_{\rho\tau\delta\xi} p^\rho_3 \epsilon^\tau_{\phi}(p^\delta_1+q^\delta)\Big]\Big[\frac{g_{X^{[D^*_s D_s]}}}{\sqrt{2}}\epsilon^{Z_{cs}}_\rho\Big]\\
			&\times \frac{-g^{\mu\xi}+p^{\mu}_1 p^{\xi}_1 /m^2_1}{p^2_1-m^2_1} \frac{-g^{\nu\rho}+p^{\nu}_2 p^{\rho}_2 /m^2_2}{p^2_2-m^2_2}\frac{1}{q^2-m^2_q}\mathcal{F}(q^2,m_q^2) \, , \\
		\end{split}
	\end{equation*}

	\begin{equation*}
	\begin{split}
		\mathcal{M}_{D^{+}_s D^{-}_s D^{*+}_s}&= i^3 \int\frac{d^4 q}{(2\pi)^4}\mathcal{A}^{B_s\rightarrow D_s \bar{D}_s}(p_1,p_2)\Big[2f_{D_s D^*_s \phi}\varepsilon_{\mu\nu\alpha\beta}p^\mu_3 \epsilon^\nu_{\phi}(p_1+q)^\alpha\Big]\Big[\frac{g_{X^{[D^*_s D_s]}}}{\sqrt{2}}\epsilon^{Z_{cs}}_\sigma\Big] \\
		&\times \frac{1}{p^2_1-m^2_1} \frac{1}{p^2_2-m^2_2}\frac{-g^{\beta\sigma}+q^{\beta} q^{\sigma} /m^2_q}{q^2-m^2_q}\mathcal{F}(q^2,m_q^2) \, ,\\
		\mathcal{M}_{D^{+}_s D^{*-}_s D^{*+}_s}&= i^3 \int\frac{d^4 q}{(2\pi)^4}\mathcal{A}_\mu^{B_s\rightarrow D_s \bar{D}^*_s}(p_1,p_2)\Big[2f_{D_s D^*_s \phi}\varepsilon_{\rho\nu\alpha\beta}p^\rho_3 \epsilon^\nu_{\phi}(p_1+q)^\alpha\Big]\Big[-\frac{g_{X^{[D^*_s D^*_s]}}}{\sqrt{2}}\varepsilon_{\omega\sigma\tau\xi}p^\omega_4 \epsilon_{Z_{cs}}^\sigma\Big] \\
		&\times  \frac{1}{p^2_1-m^2_1}\frac{-g^{\mu\xi}+p^{\mu}_2 p^{\xi}_2 /m^2_2}{p^2_2-m^2_2}\frac{-g^{\beta\tau}+q^{\beta} q^{\tau} /m^2_q}{q^2-m^2_q}  \mathcal{F}(q^2,m_q^2) \, ,\\
		\mathcal{M}_{D^{*+}_s D^{-}_s D^{*+}_s}&= i^3 \int\frac{d^4 q}{(2\pi)^4}\mathcal{A}_\mu^{B_s\rightarrow D^*_s \bar{D}_s}(p_1,p_2)\Big[g_{D^\ast_s D^\ast_s \phi} g_{\tau\theta} (p_{1\kappa}+q_{\kappa})\epsilon^\kappa_{\phi}+4f_{D^\ast_s D^\ast_s \phi}g_{\tau\kappa}g_{\theta\nu}(p^\nu_3 \epsilon^\kappa_{\phi}-p^\kappa_3 \epsilon^\nu_{\phi})\Big]\\
		&\times \Big[\frac{g_{X^{[D^*_s D_s]}}}{\sqrt{2}}\epsilon^{Z_{cs}}_\sigma\Big]\frac{-g^{\mu\tau}+p^{\mu}_1 p^{\tau}_1 /m^2_1}{p^2_1-m^2_1}\frac{1}{p^2_2-m^2_2} \frac{-g^{\theta\sigma}+q^{\theta} q^{\sigma} /m^2_q}{q^2-m^2_q}\mathcal{F}(q^2,m_q^2) \, ,\\
		\mathcal{M}_{D^{*+}_s D^{*-}_s D^{*+}_s}&=i^3 \int\frac{d^4 q}{(2\pi)^4}\mathcal{A}_{\mu\nu}^{B_s\rightarrow D^*_s \bar{D}^*_s}(p_1,p_2)\Big[g_{D^\ast_s D^\ast_s \phi} g_{\tau\theta} (p_{1\kappa}+q_{\kappa})\epsilon^\kappa_{\phi}+4f_{D^\ast_s D^\ast_s \phi}g_{\tau\kappa}g_{\theta\nu}(p^\nu_3 \epsilon^\kappa_{\phi}-p^\kappa_3 \epsilon^\nu_{\phi} )\Big]\\
		&\times\Big[-\frac{g_{X^{[D^*_s D^*_s]}}}{\sqrt{2}}\varepsilon_{\omega\sigma\lambda\xi}p^\omega_4 \epsilon_{Z_{cs}}^\sigma\Big]\frac{-g^{\mu\tau}+p^{\mu}_1 p^{\tau}_1 /m^2_1}{p^2_1-m^2_1}  \frac{-g^{\nu\xi}+p^{\nu}_2 p^{\xi}_2 /m^2_2}{p^2_2-m^2_2}\frac{-g^{\theta\lambda}+q^{\theta} q^{\lambda} /m^2_q}{q^2-m^2_q}\mathcal{F}(q^2,m_q^2)
	\end{split}
\end{equation*}

\end{document}